\documentstyle[twocolumn,prd,aps]{revtex}

\begin{document}
\draft
\title{\bf Rigid Singularity Theorem in Globally Hyperbolic Spacetimes}
\author{Makoto Narita}
\address{Department of Physics, Rikkyo University, 
Nishi-ikebukuro 3-34-1, Toshimaku, Tokyo, 171-8501  Japan
\\{\it E-mail: narita@se.rikkyo.ac.jp}
\\PACS number(s): $04.20.D_{W}$ $04.20.G_{Z}$}
\maketitle

\begin{abstract}
We show the rigid singularity theorem, 
that is, 
a globally hyperbolic spacetime satisfying the strong energy condition and 
containing past trapped sets, 
either is timelike geodesically incomplete 
or splits isometrically as space $\times$ time.  
This result is related to Yau's Lorentzian splitting conjecture.  
\end{abstract} 

\bigskip

30 years ago, Penrose-Hawking have shown that 
spacetimes are geodesically incomplete 
under some physically reasonable conditions
\cite{penrose}\cite{hawking}\cite{hawpen}\cite{hawkingellis}.  
The generic condition 
is the key assumption to induce singularities {\it rigidly}.  
Geroch improved these theorems with ``no observer horizon'' condition 
in place of the generic condition 
for the spatially closed universe
\cite{geroch66,geroch70}.  
Here, the ``no observer horizon'' condition is that for some $p\in M$, 
the set $M\setminus[I^{+}(p)\cap I^{-}(p)]$ is compact
(This definition is what Bartnik improved 
the original Geroch's definition\cite{bartnik}).  
Under this assumption 
Geroch proved that a globally hyperbolic spacetime 
which has compact acausal hypersurfaces without edge 
(i.e. compact Cauchy surfaces) 
is geodesically incomplete or Ricci flat.  
In this statement flatness implies that the spacetime is static, 
i.e. a timelike Killing vector field is 
orthogonal to a family of spacelike hypersurface, 
(of course, there exists a static and non-flat spacetime, 
e.g. the Einstein static universe).  
Unfortunately, The Geroch's proof is incomplete since it relies on
erroneous assertions of Avez.  
The correction for the error was established by Bartnik\cite{bartnik}.  
Galloway  modified Geroch's theorem\cite{galloway7}.  
He investigated various equivalent forms of a ``no observer horizon'' 
type condition.  
Bartnik conjectured that the spacetime splits isometrically
as space $\times$ time, i.e. static 
if spacetimes, which are timelike geodesically complete and 
globally hyperbolic with compact Cauchy surfaces, 
satisfy the timelike convergence condition (the strong energy condition)
\cite{bartnik}.  
Eschenburg and Galloway proved the Bartnik's conjecture 
with the modified ``no observer horizon'' condition, 
which is $I^{-}(\gamma)\supset S$ 
with $S$ and $\gamma$ being a compact partial Cauchy surface and 
an $S$-ray\cite{eschenburg2}, 
instead of global hyperbolicity.  
Furthermore, 
Galloway and Horta proved it with the condition which is 
$I^{-}(\gamma)\cap I^{+}(\beta)\neq \emptyset$ 
with $\gamma$ and $\beta$ being future and past $S$-rays 
\cite{galloway6}.  

This result is closely related to Yau's Lorentzian splitting 
conjecture\cite{yau}.  
This conjecture is proved by Eschenburg\cite{eschenburg1}, 
Galloway\cite{galloway3}, Newman\cite{newman}
and recently Galloway and Horta\cite{galloway6}, 
which reads, 
\bigskip

{\bf Theorem 1}
{\it
A spacetime $(M,g)$, which is timelike geodesically complete 
or globally hyperbolic,
obeys the timelike convergence condition
(the strong energy condition) and 
has a complete timelike line as well, 
splits as an isometric
product $({\bf R}\times V, -dt^{2}\oplus h)$.
\hfill$\Box$
}
\bigskip

(The Lorentzian splitting theorem differs slightly from Bartnik's conjecture, 
since the former assumes the existence of a timelike line whereas 
the latter assumes the existence of 
a compact spacelike hypersurface.  
The advantage of Eschenburg and Galloway's theorem\cite{eschenburg2} is 
that 
there {\it always} exists a line under 
the modified ``no observer horizon'' condition 
if there exists a compact spacelike hypersurface.  
)

The purpose of this paper is to show 
the rigid singularity theorem in globally hyperbolic spacetimes, 
that is: 
\bigskip

{\bf Theorem 2}
{\it
Let $(M,g)$ be a spacetime which has the following properties:
(A) $(M,g)$ is globally hyperbolic, 
(B) $(M,g)$ satisfies the timelike convergence 
condition, i.e., $R_{ab}K^{a}K^{b}\ge 0$ for all timelike vector 
$K^{a}$, and   
(C) for each closed set $C\subset S$, 
$E^{-}(C)$ is compact where $S$ is a Cauchy surface. 
Then $(M,g)$ either is timelike geodesically incomplete or 
splits into the Lorentzian product of $({\bf R}, -dt^{2})$ 
and $(S,h)$, where $S$ is a smooth compact spacelike 
hypersurface and $h$ is the induced metric on $S$.  
}
\bigskip

{\it Remark.}
Condition (C) has two physical properties.  
One of them is that, with condition (A), 
the spacetime is spatially closed, 
i.e. there exist compact Cauchy surfaces 
(shown in Lemma 1).  
The other is that the spacetime has no 
past observer horizons.  
This condition is weaker than Bartnik's 
condition since 
it requires both past and future.
\hfill$\Box$\\
\bigskip

Our notation and fundamental definitions 
(such as $I^{\pm},J^{\pm},E^{\pm}$) 
are the same as those of Hawking and Ellis\cite{hawkingellis} 
and Beem et al.\cite{beem1} 
unless otherwise mentioned.

Let $M$ be a timelike geodesically complete, time-oriented Lorentzian 
manifold.  
A spacetime $M$ is said to be {\it globally hyperbolic} 
if the strong causality condition holds on $M$ 
and if for any two points $p,q\in M$, $J^{+}(p)\cap J^{-}(q)$ is compact.  
Global hyperbolicity of $M$ implies the existence of a Cauchy surface for 
$M$.  The {\it Cauchy surface} is a subset of $M$ 
which every inextendible nonspacelike curves intersect exactly once.  

In order to prove the rigid singularity theorem we need a trapped set.    
We call an achronal set $S$ as {\it future (or past) trapped} 
if $E^{+}(S)$ (or $E^{-}(S)$) is compact.
In spacetimes satisfying the strong energy condition, 
one of the following conditions implies the existence of a trapped set.  
These are the existence of 
(1) a compact achronal set without edge, 
(2) a closed trapped surface and 
(3) a point $p$ such that on every past (or future) 
null geodesic from $p$ the expansion of the null geodesics from $p$ becomes 
negative.  
This means that in the universe gravitational attraction 
by the matter surpasses  repulsion.  
To apply the nature of trapped sets to our aim, 
the next proposition is useful.  
\bigskip

{\bf Proposition 1}
{\it
(Proposition 6.3.1 of Hawking and Ellis \cite{hawkingellis})
If $S$ is a future set then $\dot{S}$ is a closed, imbedded, achronal 
three dimensional $C^{1-}$ submanifold.  
\hfill$\Box$
}
\bigskip

In order to prove our main theorem 
we need to derive following two lemmas.  

\bigskip

{\bf Lemma 1}
{\it
A globally hyperbolic spacetime $(M,g)$ contains 
a compact acausal hypersurface without edge (i.e. a compact Cauchy surface)  
if the spacetime $(M,g)$ contains a future (or past) trapped set $T$. 
}
\bigskip

{\it Proof}:
$E^{+}(T)$ is a compact since $T$ is a trapped set.  
Also $E^{+}(T)=\dot{I}^{+}(T)$ because of the spacetime being globally 
hyperbolic, 
i.e. necessarily causally simple.  
Furthermore, $\dot{I}^{+}(T)$ is edgeless since it is the full boundary 
of the future of $T$ 
and, by Proposition 1, 
$\dot{I}^{+}(T)$ is a closed, imbedded, achronal three-dimensional $C^{1-}$ 
submanifold.  
Thus, $E^{+}(T)$ is a compact achronal hypersurface without edge.  
Let $S$ be a Cauchy surface in $I^{-}(E^{+}(T))$.  
$(M,g)$ admits a past directed $C^{1}$ timelike vector field.  
Each integral curve of the field intersects $S$ and $E^{+}(T)$ at most once.  
Therefore, we can define a continuous one-to-one map $\phi :E^{+}(T)
\rightarrow S$.  
Since $E^{+}(T)$ is compact, 
its image $\phi(E^{+}(T))$ is also compact and 
homeomorphic to $E^{+}(T)$.  
Thus, $\phi(E^{+}(T))$ is a three-dimensional manifold without boundary.  
If $S$ is non-compact, 
$\phi(E^{+}(T))$ must have a boundary in $S$.  
This leads to contradiction.  
Thus, $S$ is a compact Cauchy surface.  
\hfill$\Box$
\bigskip


{\it Remark.}
Lemma 1 closely relates to corollary 1 in Galloway's 
paper~\cite{gal85}.  
He has shown that a compact connected locally achronal edgeless 
subset is a Cauchy surface if the spacetime is globally hyperbolic.  
\hfill$\Box$
\bigskip


{\bf Lemma 2}
{\it
Let $(M,g)$ be a spacetime which is timelike geodesically complete 
and satisfies the conditions (A) and (C) in Theorem 2.  
Then $(M,g)$ contains a timelike line.  
}
\bigskip

{\it Proof}:
From Lemma 1 there exists a compact Cauchy surface $S$.  
Then, by standard arguments\cite{ehrgall} one can construct a 
timelike or null line $\mu$.  Let $\mu$ be a null line and meet $S$ at $p$.  
The portion $\tilde{\mu}$ of $\mu$ to the past of $p$ is contained 
in $I^{-}(p)$ or in $E^{-}(p)$.  
If $\tilde{\mu}$ is in $I^{-}(p)$, 
the point $p$ would connect with a point on $\tilde{\mu}$ by timelike curves.  
This contradicts the statement that $\mu$ is a line.  
If $\tilde{\mu}$ is in $E^{-}(p)$, $\mu$ is past imprisoned in a compact set 
$E^{-}(p)$ as condition (C).  This contradicts the condition that the 
spacetime is strongly causal.
Thus $\mu$ must be timelike.
\hfill$\Box$
\bigskip

By above lemmas we can prove 
the rigid singularity theorem in globally hyperbolic spacetimes.  
\bigskip

{\it Proof of theorem 2}:
Suppose that the spacetime is timelike geodesically complete.
By Lemma 2, 
there exists a timelike line.
Since condition (B) holds on $(M,g)$,
by Theorem 1,
the spacetime splits
as an isometric product $({\bf R}\times S, -dt^{2}\oplus h)$.
Thus, the proof is complete.
\hfill$\Box$

\bigskip

In summary, 
we would like to comment on condition (C) which 
is one of several ``no observer horizon'' conditions.  
At present, the weakest one is Galloway and 
Horta's condition~\cite{galloway6}.  
Ours is derived from Eschenburg and Galloway's condition~\cite{eschenburg2} 
which is stronger than Galloway and Horta's one.  
Therefore, we do not yet succeed to generalize previous works.  
The advantage of ours is, however, that 
we have proved the existence of a compact Cauchy surface 
from condition (C).  
(Other authors have assumed the existence of it.  Several authors 
of them think that this assumption is perfectly natural.)  
Therefore, two assumptions, the existence of a compact 
spatial section and no observer horizon, 
are not independent each other, 
since they are directly derived from condition (C).  
This suggests that condition (C) is the prime condition to 
prove the rigid singularity theorem.  
The present study takes a step forward
as compared with the previous results
\cite{bartnik}\cite{eschenburg2}\cite{galloway6}, but we are still
far from the complete proof of Bartnik's conjecture.
To do this, probably, the energy condition will play an important role 
since we must exclude asymptotically de Sitter-like spacetimes.

\end{document}